\documentclass[prl,aps,twocolumn,showpacs,preprintnumbers,superscriptaddress]{revtex4}
\usepackage{graphicx}
\usepackage{dcolumn}
\usepackage{bm}
\usepackage{amssymb}
\usepackage{pifont}
\usepackage{amsmath}
\usepackage{times}
\usepackage[nooneline]{subfigure}

\begin{document}

\title{Avalanche dynamics in fluid imbibition near the depinning transition}

\author{Marc Pradas}
\affiliation{Departament d'Estructura i Constituents de la Mat{\`e}ria, 
Universitat de Barcelona, Avinguda Diagonal 647, E-08028 Barcelona, Spain}

\author{Juan M. L{\'o}pez}
\affiliation{Instituto de F\'{\i}sica de Cantabria (IFCA), CSIC--UC, 
E-39005 Santander, Spain}

\author{A. Hern{\'a}ndez-Machado}
\affiliation{Departament d'Estructura i Constituents de la Mat{\`e}ria, 
Universitat de Barcelona, Avinguda Diagonal 647, E-08028 Barcelona, Spain}

\date{\today}

\begin{abstract}
We study avalanche dynamics and local activity of forced-flow imbibition fronts in disordered media. We
focus on the front dynamics as the mean velocity $\overline{v}$ of the interface is decreased and the pinning state is approached. Scaling arguments allow us to obtain the statistics of avalanche sizes and durations, which become power-law distributed due to the existence of a critical point at $\overline{v}= 0$. Results are compared with phase-field numerical simulations.
\end{abstract}

\pacs{05.40.-a, 47.56.+r, 64.60.Ht, 68.35.Ct}
\maketitle

In many physical systems, the response to a slow external driving usually involves
avalanches or bursts. Different examples are found in fracture
cracks~\cite{Maloy.etal_PRL2008}, granular material~\cite{Daer.etal_Nature1999},
earthquakes~\cite{Fisher_PhysRep1998}, or during imbibition of fluids in porous
media~\cite{Rost.etal_PRL2007}, among others. A particularly interesting problem in
this context is the dynamics of fronts during imbibition of fluids in porous
media~\cite{Rost.etal_PRL2007}, with its many engineering applications in fluidics and
oil-recovery technology.

Imbibition in disordered media occurs when a viscous fluid, which wets the medium
preferentially, displaces a less viscous fluid (typically air) and therefore, at
relatively low injection rates, stable fronts separating the two phases are formed (for a
recent review on imbibition see Ref.~\cite{Alava.etal_AP2004}). In the case of forced-flow
imbibition, the spatially averaged velocity of the liquid-air interface $\overline{v}$ is
kept constant by means of a constant injection rate. Then, for relatively low velocities,
the invading fluid advances in the form of spatially localized events or {\em avalanches},
as occurs in other disordered systems. 

Avalanche dynamics in imbibition is expected to be responsible for the front velocity
fluctuations. Rost {\it et al.}~\cite{Rost.etal_PRL2007} have recently shown that in the
case of imbibition, which is a locally conserved process, velocity fluctuations are
controlled by a length scale $\xi_\times$ arising from fluid conservation. This
characteristic length scale introduces a natural cutoff in the distribution of avalanche
sizes and durations, which leads to non-critical avalanche distributions and is ultimately
responsible for the lack of correlated fluctuations at large distances. The scaling
behavior of the velocity fluctuations can then be derived making use of the central limit
theorem~\cite{Rost.etal_PRL2007}. Forced-flow imbibition described by avalanches with a
fixed cutoff size has  been experimentally observed in the recent work by Planet {\it et
al.}~\cite{Planet.etal_U2008}, by means of analyzing the global velocity time series $\overline{v}(t)$.

\begin{figure}
\includegraphics[width=0.42\textwidth]{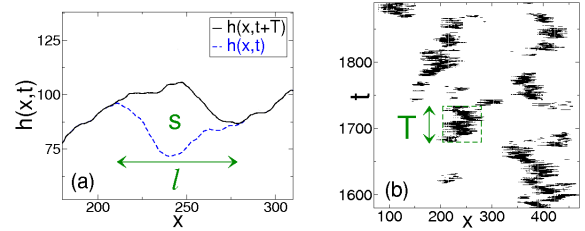}%
\caption{(Color online) (a) Typical avalanches of the front $h(x,t)$ from our phase-field simulations for $\bar{v}=v_0/40$ and clip $c_\mathrm{th}=3$. (b) Spatio-temporal activity is characterized by the size $s$, lateral
extent $\ell$, and duration $T$ of avalanches.} \label{fig:activity}
\end{figure}

In this paper we study the statistics of local avalanches of activity in forced-flow
imbibition in disordered media. We analyze the mesoscopic behavior of the interface by
monitoring locally active sites, {\it i.e.} those sites that are moving at a given time,
which define the actual avalanche taking place in the system (see
Fig.~\ref{fig:activity}). We show that avalanche sizes and durations become power-law
distributed for low enough injection rates due to the existence of a critical point at
$\overline{v}=0$. The singularity appearing as $\overline{v} \to 0$ affects the value of
the critical exponents that characterize the front dynamics and morphology-- namely, the
avalanche exponents, as well as the roughness exponents. This leads to {\em effective}
exponents even for finite velocities. We obtain scaling relations connecting the roughness
exponents and the avalanche exponents. Our scaling theory is compared with numerical
results of a phase-field model for imbibition.

Only in the last few years it has been possible to achieve a satisfactory theoretical
understanding of imbibition, based on a detailed description of the physical forces that
play a relevant role at different spatial scales~\cite{Alava.etal_AP2004}. Surface tension
$\sigma$ tends to flatten the front at short length scales, while quenched disorder in
both, capillary $p_{c}(\bm r)$  and permeability $K(\bm r)$, makes the front to roughen
and fluctuate around its average position. Both disorders operate at very different length
scales, separated by a crossover length that depends on the velocity as $\xi_{K}\sim
1/\overline{v}$~\cite{Alava.etal_AP2004,Paune.Casademunt_PRL2003}. We are concerned here
with the interesting case of a slowly advancing front, {\it i.e.} the capillary dominated
regime, where the permeability disorder is irrelevant and can be considered to be a
constant $K(\bm r)\sim K_{0}$. Different theoretical
approaches~\cite{Paune.Casademunt_PRL2003,Dube.etal_PRL1999,Hernandez-Machado.etal_EL2001}
have arrived at the conclusion that, for small deviations around the mean interface
position, the dynamical evolution for the liquid-air interface is described in Fourier
space as
\begin{equation}\label{eq:linear_h}
\partial_t{\hat{h}}_{k}=-\sigma K_0\vert k\vert k^{2}\hat{h}_{k}- \overline{v} \vert
k\vert\hat{h}_{k}+K_0\vert k\vert\hat{\eta}_{k},
\end{equation}
where $\hat\eta_{k}(\hat{\mathbf h})$ are the Fourier components of the capillary disorder
at some coarse-grained scale. From Eq.~(\ref{eq:linear_h}) one can easily see that there
exists a crossover length
\begin{equation}\label{eq:CrossoverL}
\xi_{\times}\sim \left(\frac{\sigma K_0}{\overline{v}}\right)^{1/2},
\end{equation}
such that interface fluctuations are uncorrelated above this typical scale. Indeed,
several numerical
studies~\cite{Dube.etal_PRL1999,Laurila.etal_EPJB2005,Pradas.Hernandez-Machado_PRE2006}
have shown that the interface is asymptotically flat on length scales larger than
$\xi_{\times}$, introducing then a natural cutoff in the system. For capillary-induced
fluctuations we have $\xi_{\times} \ll \xi_K$, so that the permeability disorder can be
ignored.

\paragraph{Avalanche statistics.-} In order to monitor local avalanches of forward
movements we proceed as follows. First, we define the \emph{active sites} on the interface
as those where the local velocity $v(x,t)\equiv \partial_t h(x,t)$ takes values above some
fixed threshold, $v(x,t) > c_\mathrm{th}\overline{v}$, where $c_\mathrm{th}$ is some
arbitrary constant and $\overline{v}$ is the spatially averaged global velocity. An
avalanche is defined as a connected cluster of active [$v(x,t) >
c_\mathrm{th}\overline{v}$] sites surrounded by non-active [$v(x,t) <
c_\mathrm{th}\overline{v}$] sites (see Fig.~\ref{fig:activity}).

Avalanches exhibit a typical size (volume) $\langle s (\ell) \rangle$ for an event of
lateral spatial extent $\ell$. For a given front velocity $\overline{v}$ we expect the
average avalanche size to scale with the lateral extent up to the cutoff length scale,
$\langle s(\ell) \rangle \sim \ell^{D}$ for $\ell \ll \xi_{\times}\sim
\overline{v}^{-1/2}$, where $D$ is the avalanche dimension exponent that can be easily
related with the {\em local} roughness exponent $\alpha_\mathrm{loc}$ via the local width
of the interface fluctuations $w(\ell)$. One has $\langle s(\ell) \rangle \sim l^d w(\ell)
\sim \ell^d \ell^{\alpha_\mathrm{loc}}$, and $D= d + \alpha_\mathrm{loc}$ in $d+1$
dimensions. In particular, for $d=1$, one observes that $\alpha_\mathrm{loc} =
1$~\cite{Dube.etal_PRL1999,Laurila.etal_EPJB2005,Hernandez-Machado.etal_EL2001,
Soriano.etal_PRL2005} and then $D=2$. We also expect to observe a scaling relation
$\langle s(T) \rangle \sim T^{\gamma}$ between an avalanche of duration $T$ and its size
below a certain time cutoff $T_\times$. 

\begin{figure}
\includegraphics[width=0.38\textwidth]{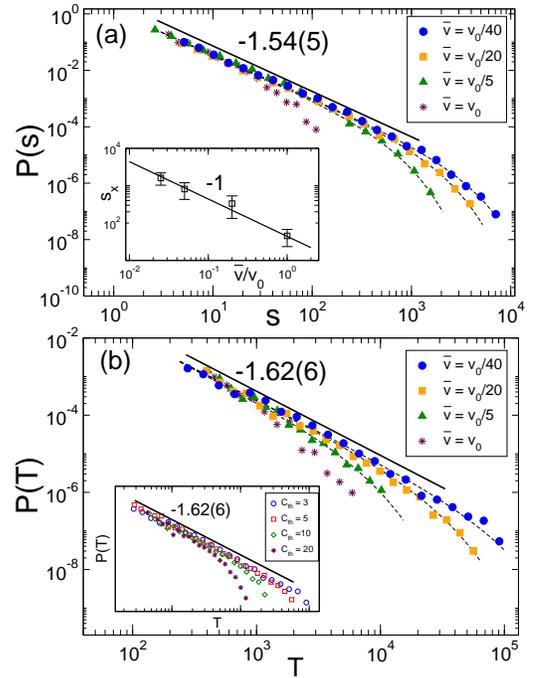}%
\caption{(Color online) (a) Distribution of avalanche size for different velocities. The
dashed curves correspond to a data fit to $\mathcal{P}(s)=a \,
s^{-\tau_s}\exp{[-s/s_{\times}(\overline{v})]}$. The solid line fits the power-law regime
for the smallest velocity. The inset shows the cutoff $s_{\times}$ of the avalanche size
distribution for each velocity with a guide-to-eye line of slope $-1$. (b) Distribution of
avalanche durations. The inset shows a comparison for different
choices of the arbitrary threshold $c_\mathrm{th}$ from $3$ to $20$ in the case of
$\overline{v} = v_0/40 = 5 \times 10^{-5}$ a.\! u.}\label{fig:distrib}
\end{figure}

In order to study the statistics of the avalanche dynamics, we first calculate the
probability densities ${\cal P}(s)$ and ${\cal P} (T)$ for having avalanches of size $s$
and duration $T$, respectively. Due to the existence of the intrinsic crossover length in
the imbibition problem the several avalanche probability distributions are not generically
expected to be critical, but exponentially decaying functions:
\begin{equation}\label{eq:distrib}
\mathcal{P}(\varrho) \sim \varrho^{-\tau_{\varrho}}\exp{-(\varrho / \varrho_\times)},
\end{equation}
where ${\cal P}(\varrho)$ is the marginal probability density function (PDF); the index $\varrho$ denotes the size
$s$, lateral extent $\ell$, or duration $T$ of avalanches, and $\tau_\varrho$ is an
exponent. The distribution cutoff depends explicitly on $\overline{v}$ and, in particular,
the maximum avalanche size is given by
\begin{equation}\label{eq:s_co}
s_\times \sim \xi_\times^D \sim (\overline{v})^{-D/2}.
\end{equation}
This cutoff diverges ($\xi_\times \to \infty$) as the control parameter $\overline{v} \to
0$, which renders critical avalanches expanding over the whole system. This divergence is
very strong, $s_\times \sim (\overline{v})^{-1}$, already in $d=1$, which -in turn- is
expected to be reflected in long-tailed avalanche distributions even for finite values of
$\overline{v}$ (see Fig. \ref{fig:distrib}). 

Let us now consider the joint probability $\mathcal{P}(s,\ell,T)$ for having an avalanche
of size $s$, extent $\ell$, and duration $T$. At the critical point $\overline{v} = 0$
scale-invariant behavior implies $\mathcal{P}(s,\ell,T)=b^{\sigma} \mathcal{P}(b^{D}s, b
\ell, b^{z} T)$ for any scaling factor $b>0$. Here, $z$ corresponds to the interface
dynamic exponent. Integrating over two of the arguments one obtains the marginal PDFs and
the scaling relations
\begin{align}\label{eq:avalexp}
 &\tau_{T}=1+\frac{D}{z}(\tau_{s}-1), & \mathrm{and}&
 &\tau_{\ell}=1+z(\tau_{T}-1),
\end{align}
that must be satisfied in the case of scale-invariant avalanche dynamics. They connect the
avalanche activity exponents ($\tau_T$, $\tau_s$, $\tau_\ell$) with the dynamics of the
front ($z$). Note that in this limit long-range interface correlations fully coincide with
avalanches of correlated events. This basically means that an avalanche occupies a
significant fraction of lateral extent of the system and cooperative correlated motion
over large scales does occur. Accordingly, close to the pinning critical point we have
$\langle \ell \rangle\sim T^{1/z}$. These scaling relations immediately imply $\langle s
\rangle \sim T^{D/z}$, so the exponent $\gamma$ relating avalanche sizes versus durations
becomes
\begin{equation}\label{eq:gamma_ours}
\gamma = D/z = (d + \alpha_\mathrm{loc})/z,
\end{equation}
in the limit $\overline{v} \to 0$.

This result is to be contrasted with the scaling relation obtained in a recent work by
Rost {\it et al.}~\cite{Rost.etal_PRL2007}. They analyzed a regime of relatively high
velocities for which the length scale $\xi_\times$ is very small as compared with system
size $L$. In this regime avalanches are very narrow and one can decompose the front motion
in independent, spatially localized, avalanches of forward moves,
$\overline{v}_\mathrm{ava} \sim (\overline{h}/\xi_\times)^d \, \overline{v}$. The
avalanche duration is $T \sim w(\ell)/v(\ell)$, where $v(\ell)$ is the front velocity over
the region of size $\ell$ spanned by the avalanche. If simultaneous avalanches are narrow
and independent events a central limit theorem argument gives $v(\ell) \sim \ell^{-d/2}$
and this leads to $\langle s(T) \rangle \sim T^\gamma$, with~\cite{Rost.etal_PRL2007}
\begin{equation}
\gamma = (\alpha_\mathrm{loc} + d)/(\alpha_\mathrm{loc} + d/2),
\label{eq:gamma_Rost}
\end{equation}
in $d+1$ dimensions. In particular, for $d=1$ one has the prediction $\gamma =
4/3$~\cite{Rost.etal_PRL2007}. Interestingly, this argument also leads to the scaling
relation $\langle \ell(T) \rangle \sim T^{\delta}$, with an exponent $\delta =
1/(\alpha_\mathrm{loc} + d/2)$ that differs from $1/z$, where $z=3$ is the dynamic
exponent describing the correlation spreading of interfacial fluctuations for forced-flow
imbibition in
$d=1$~\cite{Hernandez-Machado.etal_EL2001,Laurila.etal_EPJB2005,
Pradas.Hernandez-Machado_PRE2006,Soriano.etal_PRL2005}. This indicates that the
propagation of interface correlations is decoupled from the avalanche dynamics. Indeed, as
it will be shown below, the scaling theory leading to Eq.~(\ref{eq:gamma_Rost}) is valid
in a velocity regime such that the characteristic length scale is negligible as compared
with the system size $\xi_\times \ll L$. For lower  front velocities, when $\xi_\times$
becomes comparable with the system size, the exponent $\gamma$ should tend to the value
given by Eq.~(\ref{eq:gamma_ours}) instead of (\ref{eq:gamma_Rost}).

\paragraph{Scaling properties in the static limit.-} The interface scaling exponents can
be obtained by a scaling theory which is expected to be valid in the {\em static} (pinned
state) limit $\overline{v} \to 0$. In the pinned state the velocity-dependent term in
Eq.~(\ref{eq:linear_h}) cancels and the geometric properties of the front can be described
by the balance between surface tension and capillary disorder, which in real space can be
written as $\sigma \nabla^{2}h_\mathrm{p}(x) + \eta(x) \simeq 0$, where $h_\mathrm{p}(x)$
is the pinned state, and the disorder is delta correlated, $\langle \eta(x)\eta(x')\rangle=\eta_0^2+\Delta^2\delta(x-x')$, with a mean value $\langle\eta(x)\rangle\equiv \eta_0$ and variance $\Delta^2-\eta_0^2$. Applying a scaling transformation, $x \to b \, x$ and $h_\mathrm{p}
\to b^{\alpha} \, h_\mathrm{p}$, scale-invariance holds for a global roughness exponent
$\alpha=3/2$ for the pinned state configuration. Small perturbations of the pinned state
$\delta h$ are assumed to relax towards another of the infinitely many pinned
configurations according to
\begin{equation}\label{eq:relax}
 \partial_t (\delta h) = \sigma K_0 \nabla^{2}(\delta h) + K_0 \eta(x),
\end{equation}
which leads to the exact interface exponents $z=2$, $\alpha = 3/2$, and
$\alpha_\mathrm{loc} = 1$ at the critical point $\overline{v}=0$ for $d=1$. These
exponents can now be replaced in the avalanche scaling relations~(\ref{eq:avalexp}) and
(\ref{eq:gamma_ours}) to obtain $\tau_T = \tau_s$, $\tau_\ell = 2 \tau_T -1$, and
$\gamma=1$, where we have used $D = d + \alpha_\mathrm{loc} = 2$ in $d=1$. In the
following we compare these scaling results with numerical integrations of a phase-field
model as one approaches the singular point $\overline{v}=0$.

\paragraph{The phase-field model.-} The scaling properties of fluid imbibition fronts can
be well described by means of a phase-field
model~\cite{Dube.etal_PRL1999,Hernandez-Machado.etal_EL2001}. A conserved field $\phi$ is
used to represent the two existing phases, taking the equilibrium values $\phi_{eq}= +1$
and $\phi_{eq}= -1$ in the liquid and air phases, respectively. The dynamics of the phase
field is controlled by a continuity equation based on a time-dependent Ginzburg-Landau
model with conserved order parameter $\partial_t\phi = \bm\nabla M \bm\nabla\mu$ where
$\mu=\delta \mathcal{F}/\delta\phi$ is the chemical potential and the free energy takes
the form $\mathcal{F}[\phi]=\int d\bm{r}\left[(\epsilon\bm\nabla\phi)^{2}/2) - \phi^2/2 +
\phi^4/4 - \eta(\bm{r})\phi\right]$. The quenched random field $\eta(\bm{r})>0$ models
capillary disorder and favors the liquid (wet) phase, forcing the interface to advance at
the expense of the air (dry) phase. In our numerical model we have used a spatially
distributed dichotomic quenched noise in a two-dimensional system. The locally conserved
dynamics is described by
\begin{align}\label{eq:Phase-Field}
\partial_t\phi=\bm{\nabla}
M\bm{\nabla}\big[-\phi+\phi^{3}-\epsilon^{2}\bm\nabla^{2}\phi-\eta(\bm{r})\big],
\end{align}
where $M$ is a mobility parameter which we take constant at the liquid phase ($\phi>0$)
and zero at the air phase ($\phi<0$), and the disorder is Gaussian with a correlator $\langle\eta(\bm{r})\eta(\bm{r'})\rangle=\langle\eta\rangle^2+\langle\eta^2\rangle\delta(x-x')$. Equation (\ref{eq:Phase-Field}) is then integrated in the "weak" disorder case~\cite{Laurila.etal_PRE2008}, i.e., when the disorder intensity is much smaller than the dimensionless surface tension, in a system size of $L=512$ and 25 disorder realizations with $\epsilon=1$, and the forced-flow boundary condition  
$\bm{\nabla}\mu=-\bar{v}\hat{y}$ is imposed at the bottom of the system~\cite{Laurila.etal_EPJB2005}. All the values for the average front velocity have
been normalized to a reference value $v_{0}=0.002$, which corresponds to the highest value
studied in this paper.
\begin{figure}
\includegraphics[width=0.38\textwidth]{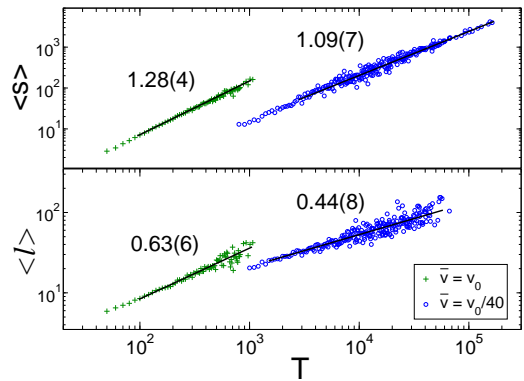}%
\caption{(Color online) Average avalanche size and lateral extent versus duration in a system of size 	$L=512$
for two velocities in the different dynamical regimes discussed in this paper.}
\label{fig:lst}
\end{figure}

\paragraph{Numerical results.-} 
Figure~\ref{fig:distrib} shows the avalanche size and
duration statistics calculated using a threshold $c_\mathrm{th}=3$. We observe that the probability distributions tend to a power-law as
$\overline{v}$ is decreased, due to the divergent cutoff [cf. Eq.~(\ref{eq:s_co})]. We
estimate the exponents $\tau_{s}\simeq 1.54(5)$ and $\tau_{T}\simeq 1.62(6)$ from the
scaling of the data. Despite the smallest velocity we were able to reach $\overline{v} =
v_0/40 = 5 \times 10^{-5}$, which is still far from zero, the scaling region is reasonably good.
According to our scaling theory both exponents should exactly coincide at the critical
point, which is consistent with the numerical values within the error bars. We also plot a
direct estimate of the divergent avalanche size cutoff in good agreement with
Eq.~(\ref{eq:s_co}).

We also find an excellent agreement with our prediction in Eq.~(\ref{eq:gamma_ours}) for
the scaling relation between size and duration of an avalanche, $\langle s(T) \rangle \sim
T^{\gamma}$. In Fig.~\ref{fig:lst} we plot both avalanche size and lateral extent {\it vs}
time for two typical velocities. For the lowest velocity we studied $\overline{v} = v_0/40
= 5 \times 10^{-5}$ we estimate $\gamma \simeq 1.09 (7)$, which is to be compared with
$\gamma = 1$ from Eq.~(\ref{eq:gamma_ours}) in $d=1$. This can also be compared with the
scaling relations between the avalanche exponents $\tau_s$ and $\tau_T$ given by
Eq.~(\ref{eq:avalexp}). Substituting the numerical values $\gamma = 1.09$ and $\tau_T =
1.54$ we predict $\tau_t\simeq 1.60$ in good agreement with the numerical result
(cf.~Fig.~\ref{fig:distrib}). Note that the scaling theory is expected to be exact only at
the critical point $\overline{v}=0$, which is not actually reached with our phase-field
model results. However, the singularity is strong enough to lead to effective exponents
for velocities within a critical region $\overline{v} \leq L^{-2}$.

At variance with Rost {\it et al.}~\cite{Rost.etal_PRL2007}, who only monitored avalanches
in the global velocity time series $\overline{v}(t)$, here we are actually looking at
active sites that participate in an avalanche and, therefore, we are able to check the
validity of the scaling law $\langle \ell \rangle \sim T^\delta$. The typical lateral
extent is predicted to scale with avalanche duration with an exponent $\delta =
1/(\alpha_\mathrm{loc}+ d/2) = 2/3$ in the high velocity regime for
$d=1$~\cite{Rost.etal_PRL2007}, in excellent agreement with our numerical estimate in
Fig.~\ref{fig:lst}. However, for low velocities we predict $\delta = {1/z} = 1/2$ with the
dynamic exponent $z =2$ in the static limit. A strong proof of a distinctive behavior as
the front velocity is decreased can be readily seen in Fig.~\ref{fig:lst}. Our numerical
simulations indicate that $\gamma \to 1$ and $\langle \ell \rangle \sim T^{1/2}$ as
$\overline{v} \to 0$, pointing out that the dynamics is controlled by the static critical
point at $\overline{v} = 0$.
\begin{figure}
\includegraphics[width=0.45\textwidth]{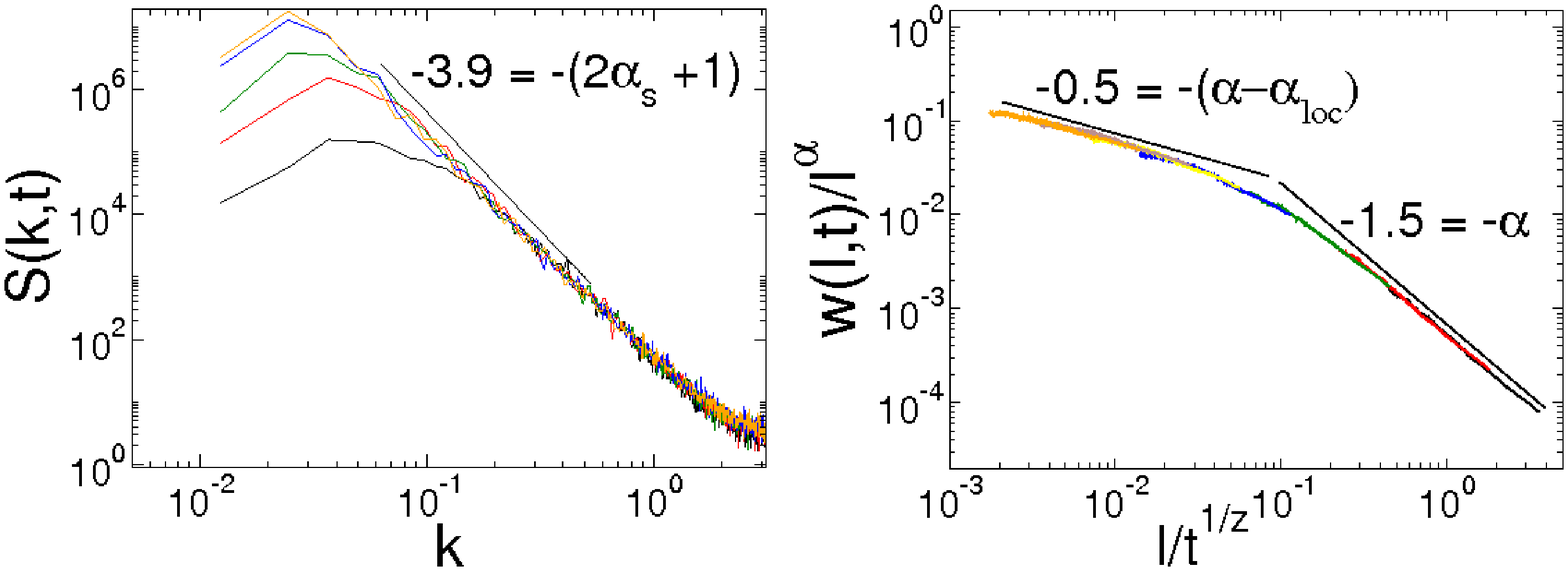}%
\caption{(Color online) Scaling of the front roughness in the phase-field simulations for $L=512$ in the low velocity $\overline{v}=v_{0}/40$ limit. (a) Structure factor $\mathcal{S}(k,t)$ at different times giving a spectral
roughness exponent of $\alpha_{s}=1.45(6)$~\cite{Ramasco.etal_PRL2000}. (b) Data collapse
of the local width according to a super-rough scaling for
$\alpha=1.50(3)$, $\alpha_\mathrm{loc} = 1.0$, and $z=2.09(6)$.}\label{fig:scaling}
\end{figure}

On the other hand, we have also estimated the scaling of the interfacial fluctuations for
states close to the static limit. This provides an independent check of the validity of
our scaling relations in Eqs.~(\ref{eq:avalexp}) and (\ref{eq:gamma_ours}) connecting
avalanche and roughness exponents. From the structure factor ${\cal
S}(k,t) = \langle |h_{k}(t)|^2 \rangle$ and local width in Fig.~\ref{fig:scaling}, we estimate the global,
local, and spectral roughness exponents~\cite{Ramasco.etal_PRL2000} $\alpha = 1.50(3)$,
$\alpha_\mathrm{loc} = 1$, $\alpha_\mathrm{s} = 1.45(6)$, respectively, and the dynamic
exponent $z = 2.09(6)$ in excellent agreement with our scaling theory for the pinned
state.
\begin{table}
\centering%
\begin{ruledtabular}
\begin{tabular}{c|cccc}%
  $\overline{v}$        & $\xi_{\times}/L$ &  $\alpha$  &  $z$  & $\gamma$ \\
\hline%
  $v_{0}$    &  0.1  &   1.33    &   3   &   1.28  \\
  $v_{0}/5$  &  0.23 &  1.35    &  2.8  &   1.21   \\
  $v_{0}/20$ &  0.48 &  1.41    &  2.3  &   1.13   \\
  $v_{0}/40$ &  0.64 &  1.50    &  2.09  &   1.09   \\
 \end{tabular}%
\end{ruledtabular}
\caption{Correlated extend and effective scaling exponents from phase-field simulations
at different velocities in as system of size $L=512$. The fronts are always super-rough with $\alpha_\mathrm{loc}=1$~\cite{Ramasco.etal_PRL2000}.
}\label{tab:scalexp}%
\end{table}

Finally, we claim that the existence of a singular behavior as $\overline{v} \to 0$ and
the extent of the critical region $\overline{v} \leq L^{-2}$ explain earlier numerical
observations~\cite{Laurila.etal_EPJB2005,Alava.etal_AP2004} that reported a dependence of
the critical exponents $\alpha(\overline{v})$ and $z(\overline{v})$ with the velocity in
numerical results of forced-flow imbibition in finite systems. Table~\ref{tab:scalexp}
summarizes the different interfacial scaling exponents we observed for different
velocities. We observe that as the static limit is approached the dynamic exponent $z \to
2$ and the roughness exponent $\alpha \to 3/2$, as corresponds to the pinned state.

To conclude, we have studied avalanche dynamics in forced-flow imbibition in the pinning 
limit $\overline{v}\to 0$. A scaling theory relating the roughness of the front with the 
avalanche dynamics has been developed in excellent agreement with numerical results. Our 
scaling analysis is based on the presence of long-range correlations due to the divergent 
characteristic length scale at $\overline{v} \to 0$.
From an experimental point of view, it would be of great interest to explore the pinning limit.
In the experimental setup of a Hele-Shaw cell~\cite{Planet.etal_U2008}, this limit may be achieved by
putting the cell at an angle so that gravity plays a
role. This setup should produce fronts near pinning.
Note that already in the case of having
 a correlation length of about a 30\% of the system size, the effect of the critical point should show
 up in a drift of the measured scaling exponents (see Table I). 
Alternatively,
fluctuations around the critical point $\overline{v}=0$ could also be experimentally
tested by setting
the cell at an angle, so that the front is pinned, and then study how the system responds
to a small angle variation. In this configuration we expect the front to jump from one
pinned state to another following a relaxation dynamics described by Eq.~(\ref{eq:relax})
driven by avalanches described by Eqs.~(\ref{eq:avalexp}) and (\ref{eq:gamma_ours}).

\acknowledgments We thank R. Planet, J. Ort\'in and S. Santucci for discussion on their
experimental work. This work is supported by the DGI (Ministerio de Educaci\'on y Ciencia,
Spain) through Grant Nos. FIS2006-12253-C06-04 and -05.

\end{document}